\documentclass[oldversion]{aa} 

\usepackage{graphicx}

\newcommand{\Msun} {M$_\odot$}

\newcommand{\Mstar} {M$_\star$}

\newcommand{\grad} {$^{\rm o}$}

\newcommand{\simless}{\mathbin{\lower 3pt\hbox 
      {$\rlap{\raise 5pt\hbox{$\char'074$}}\mathchar"7218$}}} 
\newcommand{\simgreat}{\mathbin{\lower 3pt\hbox 
     {$\rlap{\raise 5pt\hbox{$\char'076$}}\mathchar"7218$}}}  
 
\begin{document}  
 
\title{Millimeter imaging of HD~163296: probing the disk structure and
  kinematics\thanks{Based on observations carried out with IRAM
  Plateau de Bure Interferometer, Submillimeter Array and NRAO Very Large
  Array. IRAM Plateau de Bure Interferometer is supported by INSU/CNRS
  (France), MPG (Germany) and IGN (Spain).The Submillimeter Array is a
  joint project between the Smithsonian Astrophysical Observatory and
  the Academia Sinica Institute of Astronomy and Astrophysics and is
  funded by the Smithsonian Institution and the Academia Sinica. The
  NRAO is a facility of the National Science Foundation operated under
  cooperative agreement by Associated Universities, Inc.}}  
\subtitle{}

\author{
Andrea Isella \inst{1,2}
\and
Leonardo Testi \inst{1}
\and
Antonella Natta  \inst{1}
\and
Roberto Neri \inst{3}
\and
David Wilner \inst{4}
\and
Chunhua Qi \inst{4}
} 

\offprints{A. Isella, \email{ isella@arcetri.astro.it}} 

\institute{ 
  Osservatorio Astrofisico di Arcetri, INAF, Largo E.Fermi 5, 
  I-50125 Firenze, Italy 
  \and 
  Dipartimento di Fisica, Universit\'a di Milano, Via Celoria 16,
  20133 Milano, Italy  
  \and
  Institut de Radio Astronomie Millim\'etrique (IRAM), 300 Rue de la
  Piscine, Domaine Universitaire de Grenoble, St. Martin d'H\'eres
  F-38406, France
  \and
  Harvard­-Smithsonian Center for Astrophysics, 60 Garden Street, MS
  42, Cambridge, MA 02138, USA
}

\date{Received ...; accepted ...} 
 
\authorrunning{Isella A. et al.} 
\titlerunning{Millimeter imaging of HD~163296}

\abstract 
{We present new multi-wavelength millimeter interferometric
  observations of the Herbig Ae star HD~163296 obtained with the
  IRAM/PBI, SMA and VLA arrays both in continuum and in the $^{12}$CO,
  $^{13}$CO and C$^{18}$O emission lines.  Gas and dust properties
  have been obtained comparing the observations with self-consistent
  disk models for the dust and CO emission. The circumstellar
  disk is resolved both in the continuum and in CO. We find
  strong evidence that the circumstellar material is in Keplerian
  rotation around a central star of 2.6~\Msun. The disk inclination
  with respect to the line of sight is 46\grad$\pm$4\grad\ with a
  position angle of 128\grad$\pm$4\grad. 
  The slope of the dust opacity measured between 0.87 and 7~mm
  ($\beta=1$) confirms the presence of mm/cm-size grains in the disk
  midplane. The dust continuum emission is asymmetric and
  confined inside a radius of 200~AU while the CO emission extends up
  to 540~AU. The comparison between dust and CO temperature indicates
  that CO is present only in the disk interior. Finally, we
  obtain an increasing depletion of CO isotopomers from $^{12}$CO to
  $^{13}$CO and  C$^{18}$O. We argue that these results support the
  idea that the disk of HD~163296 is strongly
  evolved. In particular, we suggest that there is a strong depletion
  of dust relative to gas outside 200~AU; this may be due to the
  inward migration of large bodies that form in the outer disk or
  to clearing of a large gap in the dust distribution by a low mass
  companion.  
}

\maketitle

\section{Introduction} 
\label{sec:intro}
Millimeter and sub-millimeter interferometers are providing an
increasingly detailed description of disks around pre-main sequence
stars of solar (T Tauri stars; TTS) and intermediate mass (Herbig Ae;
HAe). Both dust continuum emission  and emission in  molecular
lines are observed and spatially resolved in a number of disks,
yielding information on the disk density and temperature, the dust
properties and the gas chemistry and dynamics  in the outer disk
(e.g., Natta et al.~\cite{Nea_PPV}, Dutrey et al.~\cite{Dea_PPV}, and
references therein). Thanks to the recent instrumental improvements,
it is now possible to build upon the original detections and 
study more accurately the disk structure details.
The number of well-studied disks is however still very small,
practically restricted to the most massive and luminous 
ones; still, it is clear that disks differ from one another. Recently,
it has been reported evidence of spiral structures in AB Aur, a 2-3
Myr old intermediate mass star, and of deviations from Keplerian
rotation (Pi\'etu et al.~\cite{Pea05}; Corder et al.~\cite{Cea05});
the classical TTS LkCa15 has  a large inner hole of size $\sim 50$ AU
depleted of dust, while the HAe star MWC~480 has a smooth disk with an
optically thick (at millimeter wavelengths) inner region of radius
$\sim 35$ AU  (Pi\'etu et al.~\cite{Pea06}). Both spiral structures
and large gaps are evidence of dynamical perturbations, possibly due
to the effect of large planets.  The existence of both unperturbed and
distorted disks among pre-main sequence stars suggests that the planet
formation is actively occurring during this evolutionary stage,
leaving detectable marks on the parent disks. It is therefore
important to study in detail as many disks as possible, in order to
characterize their basic properties and to detect deviations from the
simple patterns of homogeneous disks in Keplerian rotation. 

We report in this paper a detailed study of the disk associated to the 
HAe star HD~163296, using observations in the continuum and CO lines
obtained with three different interferometers, namely the Very Large
Array (VLA) at 7mm, IRAM Plateau de Bure Interferometer (PBI) at 1.3
and 2.6~mm and the Submillimeter Array (SMA) at 0.87~mm. HD~163296 is 
a star of spectral type A1, mass of roughly 2.3~\Msun, distance
122~pc (van den Ancker et al.~\cite{vdA98}). Early OVRO observations
(Mannings and Sargent \cite{MS97}) have shown the presence of a disk
with a minimum  mass $\sim 0.03$ \Msun\ and evidence of rotation from
the CO lines. The disk is seen in scattered light by Grady et
al.~(\cite{G2000}, \cite{Gea99}),  with radius of $\sim 500$ AU; it
has an associated jet seen in Ly-$\alpha$ with HST, extending on both
sides of the disk orthogonally to it (Devine et al.~\cite{Dev00},
Wassell et al.~\cite{W06}). Natta et al.~(\cite{Nea04}) found evidence
of evolved dust in the outer disk of HD~163296 by comparing the VLA
7~mm flux to the OVRO observations.    

The results we present here have much higher spatial resolution and
wavelength coverage than what has been previously reported. They allow us to
measure accurately the dynamics of the disk as well as the disk and
dust properties and to test the capability of disk models to account
for the observations. As we will show, they suggest that the HD~163296
system is probably evolving towards a debris disk phase. 

The structure of the paper is as follows. Sec.~\ref{sec:obs} will  
describe the observations. The results will be presented in 
Sec.~\ref{sec:obs_res}, where we will derive some of the disk
parameters. A more detailed analysis, using self-consistent disk
models of the dust and CO line emission will be presented in
Sec.~\ref{sec:model}; Sec.~\ref{sec:res} contains the results, which
will be further discussed in Sec.~\ref{sec:disc}. Summary and 
conclusions follow in Sec.~\ref{sec:sum} and \ref{sec:conc}.

\section{Observations and data reduction }
\label{sec:obs}

\subsection{PBI observations}
\label{sec:obs_PBI}

The PBI observations
were carried over the 2003/2004 winter season. The six 15~m dishes
  were used in the most extended   
configuration providing a baseline coverage between 25 and 400~m. The
corresponding angular resolutions are reported in
Tab.~\ref{tab:obs}. The receivers were tuned to observe the $^{12}$CO
J=2--1 line and the nearby continuum at 1.3 mm, while at 2.8 mm the 
$^{13}$CO J=1--0, and C$^{18}$O J=1--0 lines were observed along with 
the continuum. Bandpass and complex gain calibrations were
ensured by observations of standard IRAM calibrators. The phase
stability was excellent during our observations and only a minimal
amount of editing of the data was necessary. All calibrations were
performed using the standard CLIC suite of programmes within the
GILDAS software package. The calibrated uv data were then exported
for the subsequent analysis. The accuracy of the flux density scale  
calibration is expected to be within 20\%\ at these wavelengths.

\subsection{SMA observations}

The SMA observations of
HD~163296 were made on August 23rd, 2005 using the Compact
Configuration of seven of the 6 meter diameter antennas, which
provided 21 independent baselines ranging in length from 8 to 80
meters. The SMA digital correlator was configured with a narrow band
of 512 channels over 104 MHz, which provided 0.2 MHz frequency
resolution, or 0.18 km~s$^{-1}$ velocity resolution at 345 GHz, and
the full correlator bandwidth was 2 GHz. The weather was good with
$\tau$(225 GHz) around 0.06 and the double-sideband (DSB) system
temperature were between 200 and 500 K. The source HD~163296 was
observed from HA -3 to 4.5. Calibration of the visibility phases and
amplitudes was achieved with observations of the quasar 1921-293,
typically at intervals of 25  minutes. Observations of Uranus provided
the absolute scale for the flux density calibration and the
uncertainties in the flux scale are  estimated to be 20\%. The data
were calibrated using the MIR software
package\footnote{http://cfa-www.harvard.edu/~cqi/mircook.html}. 

\subsection{VLA observations}
\label{sec:obs_VLA}

HD~163296 was observed at the NRAO/VLA as part of a larger
  survey for 7~mm  disk emission around Herbig Ae stars (see Natta et
  al., \cite{Nea04}). Data were obtained with the array in the C and D 
configurations, in several occasions from Dec 2001 through May
2003. Accurate pointing was ensured by hourly pointing sessions at
3.6~cm on a bright extra galactic object. The array offered baselines
from the shadowing limit through $\sim 3.4$~km, although some data was
obtained in a hybrid DnA configuration, all the data from the longer
baselines had to be rejected due to large phase fluctuations that
could not be corrected. The resulting uv-plane coverage offered an
angular resolution of 0.5$\arcsec$ and a maximum recoverable size of
the order of $\sim$40$\arcsec$. All the data was edited and calibrated
using standard recipes in AIPS. The short term complex gain variations
were corrected using frequent (few minutes cycle) observations of the
quasar 1820$-$254, while the flux scale was set observing the VLA
calibrator 1331$+$305. This procedure is expected to be accurate
within $\sim 15$\%\ at 7~mm.

\section{Observational results}
\label{sec:obs_res}

\begin{figure*} 
\begin{center} 
\includegraphics[width=5cm, angle=270]{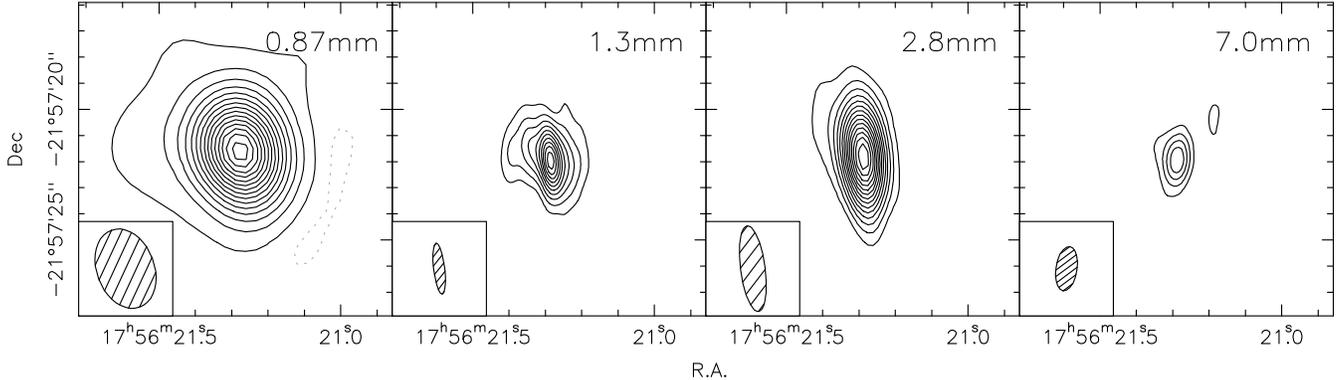} 
\caption{ \label{fig:cont}  Continuum maps of HD~163296 at 0.87, 1.3,
  2.8 and 7~mm, starting form the left. In order to highlight the
  extended  morphology of the disk, the first contour level In the
  0.87~mm map corresponds to 30~mJy (3$\sigma$), the second to
  6$\sigma$ while the inner contour levels are spaced by 10$\sigma$. At
  longer wavelength the contour levels are all spaced by 3$\sigma$,
  corresponding to 12 mJy at 1.3 
  mm, 3.3~mJy at 2.8~mm and 0.75~mJy at 7~mm. The small boxes show the
  relative synthesized beams. The integrated fluxes, the beam
  dimensions and orientations are summarized in Tab.~\ref{tab:obs}}.   
\end{center}
\end{figure*}

\begin{figure*}
\begin{center}
\includegraphics[width=18cm, angle=0]{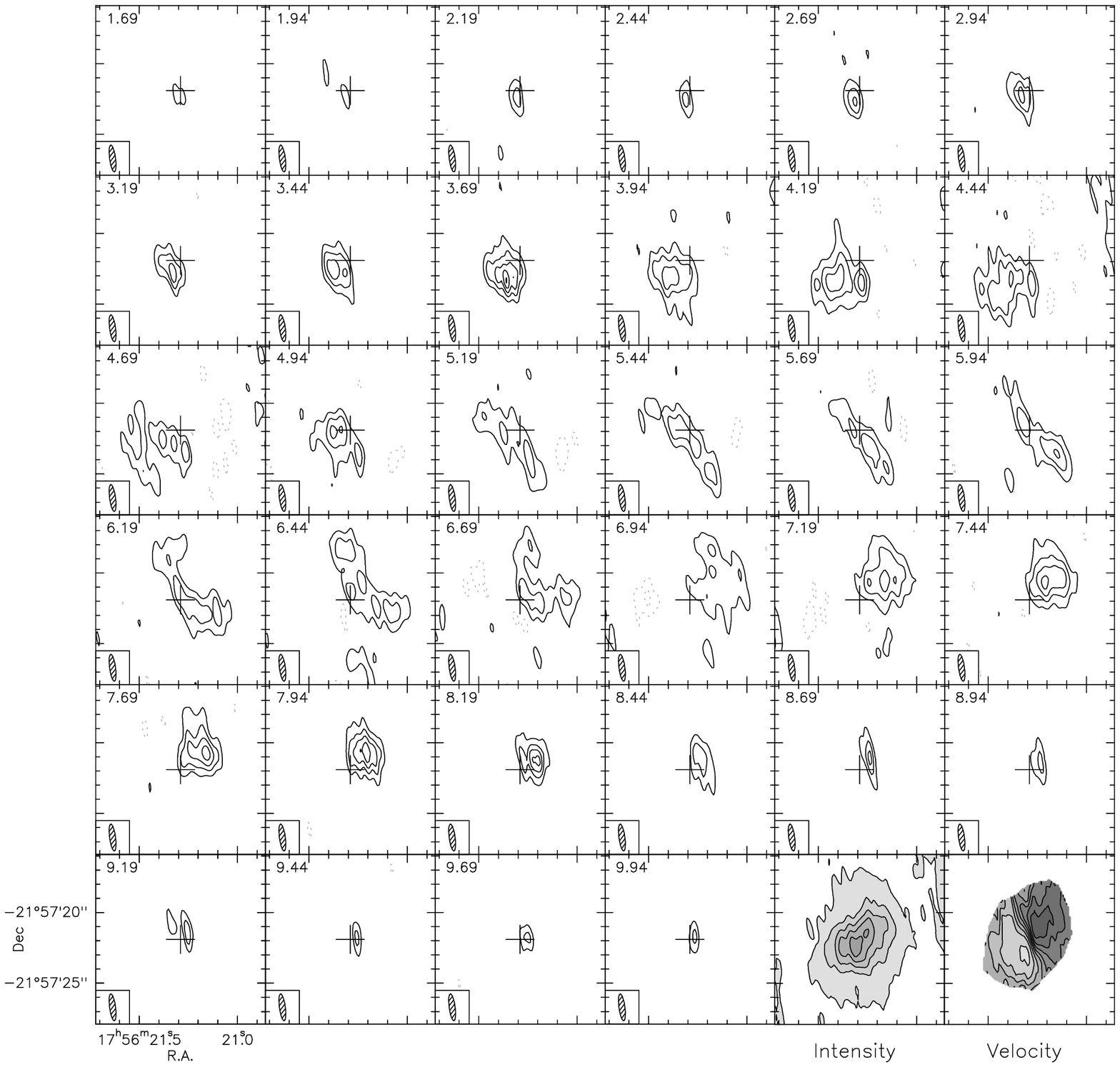} 
\caption{ \label{fig:12CO} Velocity channel maps of the $^{12}$CO
  J=2--1 emission. The velocity resolution is
  0.25~km/sec. The LSR velocity is indicated in the upper left corner
  of each panel. The angular resolution (synthesized beam), indicated
  in the small boxes, is 1.95$\arcsec \times 0.42\arcsec$ at PA
  7\grad; the contour spacing is 0.23~Jy/beam corresponding to
  3$\sigma$.  The last two panels show the integrated intensity
  (contour levels spaced by 0.4~Jy/beam km s$^{-1}$) and the velocity
  field (contour levels from 3~km~s$^{-1}$ to 9~km~s$^{-1}$ spaced by
  0.5~km~s$^{-1}$) respectively.  
  }
\end{center}
\end{figure*}

\begin{figure*}
\begin{center}
\includegraphics[width=18cm, angle=0]{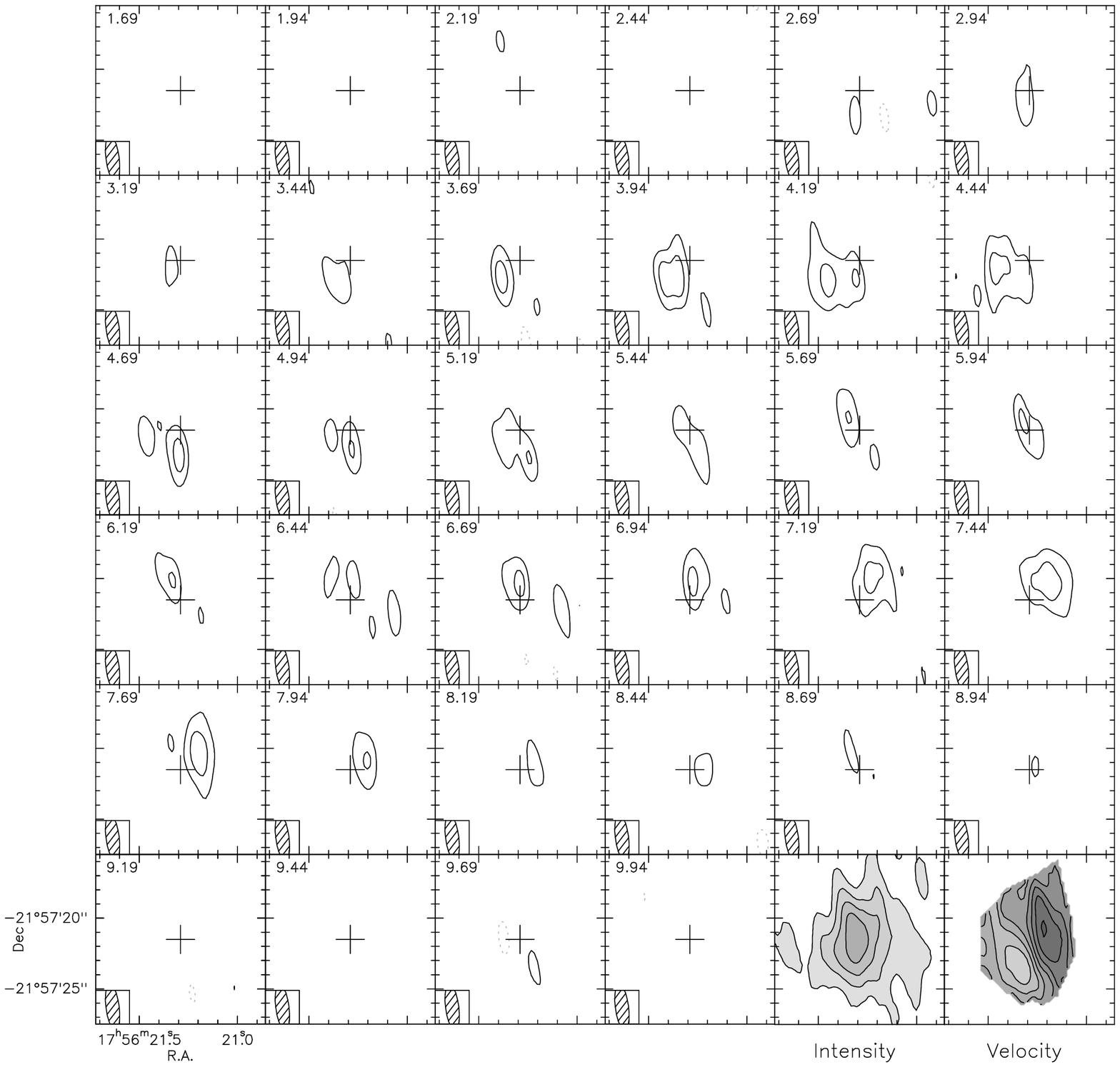} 
\caption{ \label{fig:13CO} Velocity channel maps of the $^{13}$CO
  J=1--0 emission. The velocity resolution is
  0.25~km/sec. The LSR velocity (km s$^{-1}$) is indicated in the
  upper left corner of each panel. The angular resolution (synthesized
  beam), indicated in the small boxes, is 3.3$\arcsec \times
  0.94\arcsec$ at PA 8\grad; the contour spacing is 0.09~Jy/beam
  corresponding to 3$\sigma$. The last two panels show the integrated
  intensity (contour levels spaced by 0.12~Jy/beam) and the velocity
  field (contour levels from 3~km~s$^{-1}$ to 9~km~s$^{-1}$ spaced by
  0.5 km s$^{-1}$) respectively.  
  }
\end{center}
\end{figure*}

\begin{figure*}
\begin{center}
\includegraphics[width=18cm, angle=0]{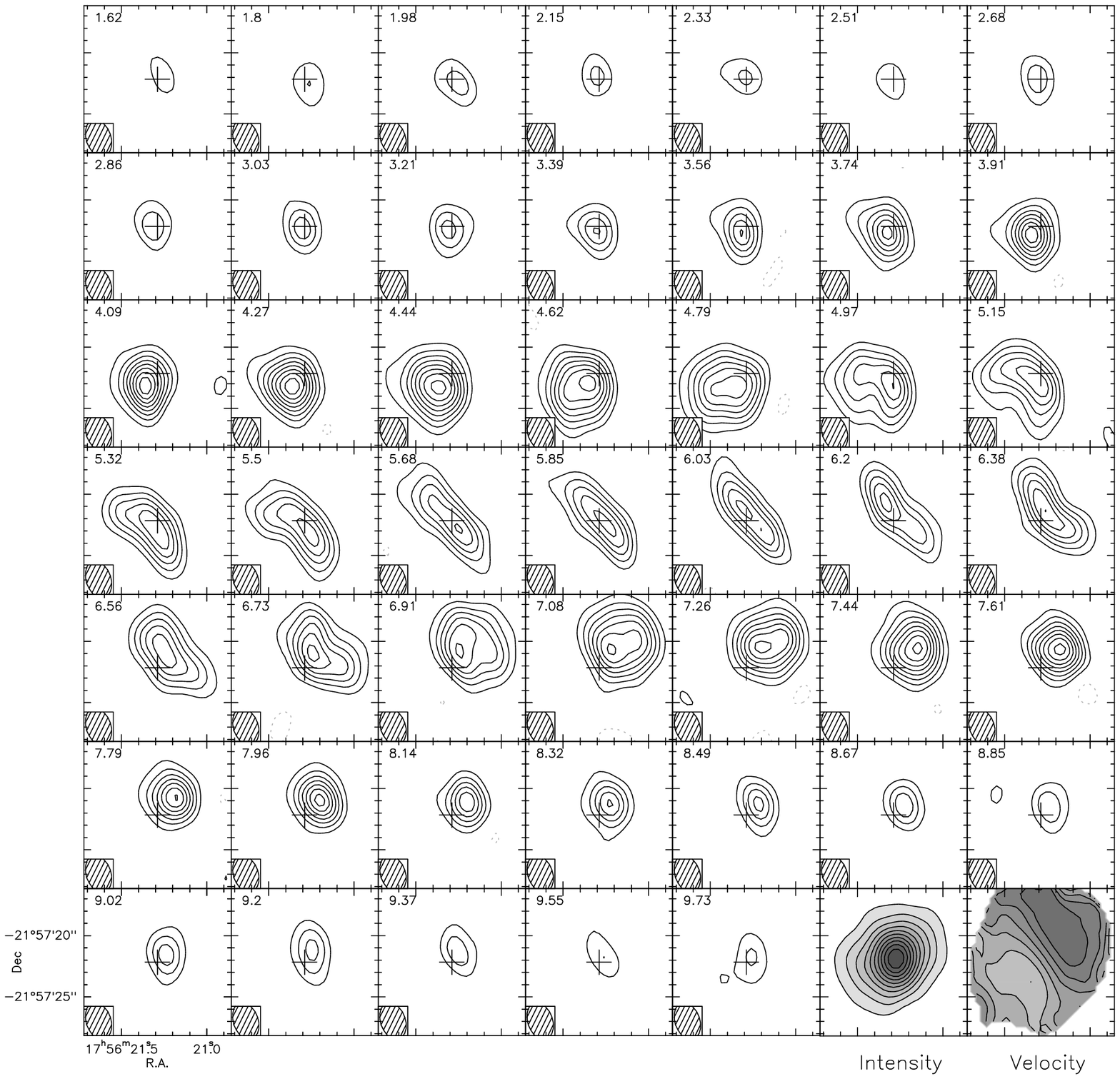} 
\caption{ \label{fig:12CO32} Velocity channel maps of the $^{12}$CO
  J=3--2 emission. The velocity resolution is
  0.18~km/sec. The LSR velocity (km s$^{-1}$) 
  is indicated in the upper left corner of each panel. The angular
  resolution (synthesized beam), indicated in the small boxes, is
  3.4$\arcsec \times 2.21\arcsec$ at PA 20\grad; the contour spacing
  is 1.5~Jy/beam corresponding to 3$\sigma$. The last two panels show
  the integrated intensity (contour levels spaced by 5~Jy/beam km
  s$^{-1}$) and the velocity field (contour levels from 3~km~s$^{-1}$
  to 9~km~s$^{-1}$ spaced by 0.5~km~s$^{-1}$) respectively. 
  }
\end{center}
\end{figure*}

The continuum maps obtained at 0.87~mm, 1.3~mm, 2.8~mm and 7~mm are
shown in Fig. \ref{fig:cont}. Channel maps, integrated intensity and
mean velocity maps of the $^{12}$CO J=2--1,  $^{13}$CO J=1--0 and
$^{12}$CO J=3--2 transitions are shown in Fig. \ref{fig:12CO},
\ref{fig:13CO} and \ref{fig:12CO32} respectively. At 
the wavelength corresponding to the C$^{18}$O J=1--0 transition we did
not detected any emission; the continuum integrated fluxes are
reported in Tab. \ref{tab:obs}.  

With simple physical assumptions,  these observations allow us to
determine some fundamental parameters of the star+disk system as the
stellar and the disk masses, the  contribution to the observed fluxes
of free-free gas emission, the wavelength dependence of the observed
flux at millimeter wavelengths  and the related dust grain opacity.  

\begin{table}
\begin{center}
\begin{tabular}{lllll}
$\lambda$ & beam & & emission & flux \\
(mm) & FWHM & PA & FWHM & (mJy)\\
\hline
\hline
0.87 & $3\arcsec.14 \times 2\arcsec.21$ & 20\grad & $3\arcsec.60
\times 2\arcsec.64$ & 1910$\pm$20 \\
1.3 & $1\arcsec.95 \times 0\arcsec.42$ &  7\grad & $2\arcsec.01 \times
0\arcsec.83$ & 705$\pm$12  \\
2.8 & $3\arcsec.3 \times 0\arcsec.94$ & 8\grad & $3\arcsec.5 \times
1\arcsec.4$ & 77.0$\pm$2.2  \\
7   & $1\arcsec.71 \times 0\arcsec.81$ & 172\grad & $1\arcsec.71
\times 0\arcsec.94$ & 4.5$\pm$0.5 \\
\hline
 \end{tabular}
\caption{\label{tab:obs}Column 2 and 3 show the size and the position angle
  of the synthesized beam of the HD~163296 observations performed with
  SMA (at 0.87~mm), PBI (at 1.3~mm and 2.7~mm) and VLA (at
  7~mm). Column 4 show the FWHM size of the detected emission along
  the major and minor axis of the beam. Column 5 shows the
  integrated flux with the uncertainties due to the statistical
  errors.}
\end{center}
\end{table}

\subsection{Disk morphology and apparent size}
\label{sec:obs_morf}

The $12\arcsec \times 12\arcsec$ continuum maps of HD~163296 are shown  
in Fig.~\ref{fig:cont}. At all wavelengths, the peak is  coincident
with the position of the optical star as measured from Hipparcos and,
given the respective synthesized beams FWHM (see Tab.~\ref{tab:obs}),
the emission is resolved and elongated approximately in the east-west
direction. Approximating the source with a circularly symmetric
geometrically thin disk and taking into account the beam shape, the 
observed aspect ratio of the level contours implies an inclination of
the disk plane from the line of sight of 45\grad~$\pm$~20\grad\
and a position angle of 120\grad~$\pm$~30\grad, in rough agreement
with the values obtained by Mannings and Sargent (\cite{MS97})
using marginally resolved OVRO observations (58\grad\ and 126\grad\
respectively).    

The emission is resolved at all the wavelengths in the East-West
direction. The fainter contour levels are
not centrally symmetric, showing excess emission in the east half of the
image, better visible in the 0.87 and 1.3~mm maps. Both the disk size
and the morphology will be discussed in more detail in
Sec.~\ref{sec:disc}.

\subsection{Disk kinematics}
\label{sec:kine}
\begin{figure}
\begin{center}
\includegraphics[width=8cm, angle=0]{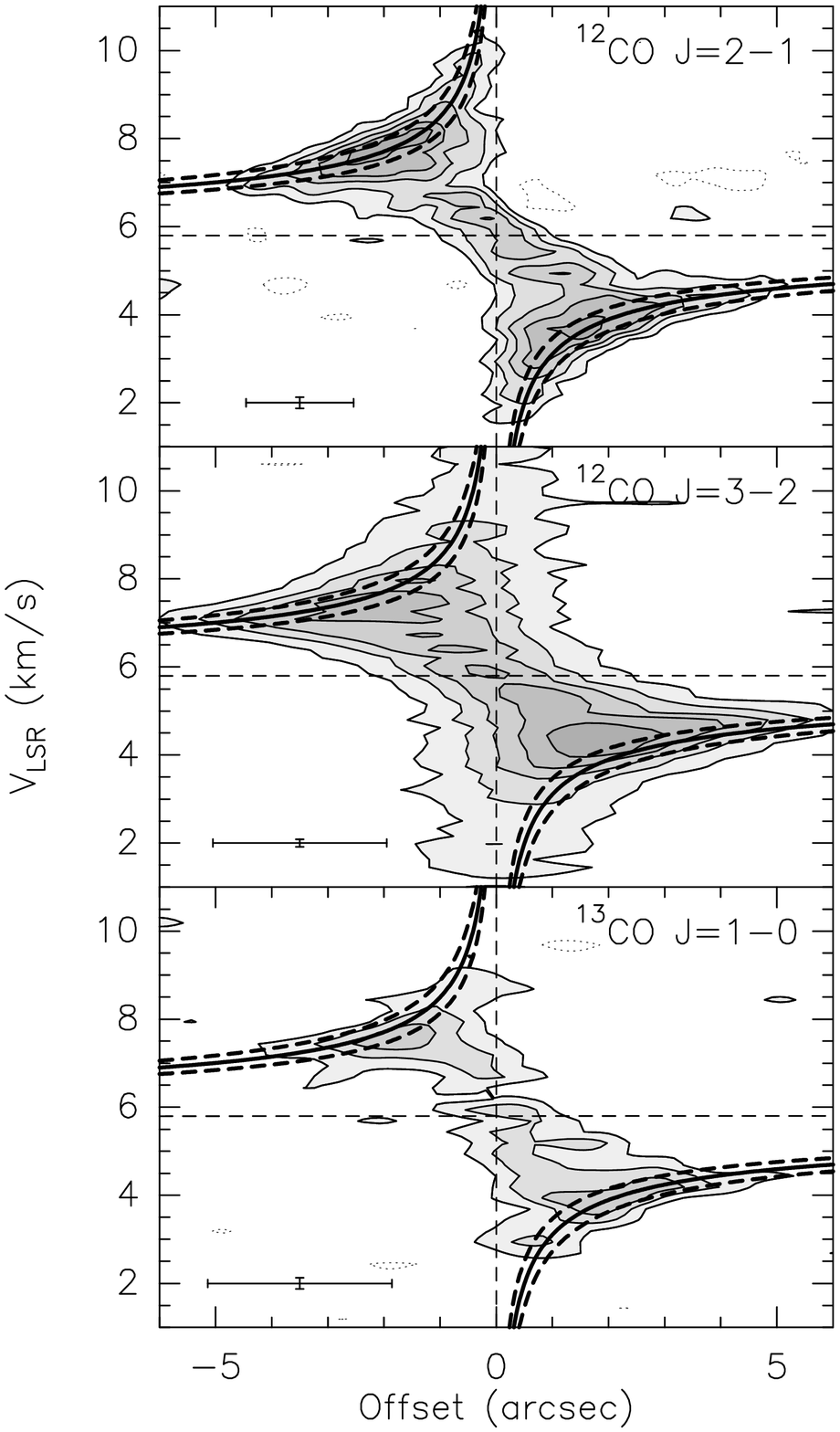} 
\caption{ \label{fig:kep_rot} Velocity-position plots along the plane
  of the disk for the $^{12}$CO J=2--1 (upper panel), the $^{12}$CO
  J=3--2 (middle panel) and the $^{13}$CO J=1--0 (lower panel)
  transitions. The angular offset is measured with respect to 
  the phase center of the observations corresponding to the position
  of the central star. The contour levels are spaced by 
  2$\sigma$ corresponding to 0.14~Jy/beam, 1~Jy/beam and 0.06~Jy/beam
  respectively. The cross in the lower  left of each panel gives the
  angular and spectral resolution of the corresponding map. The thick
  solid lines marks the border where emission is expected for a
  Keplerian disk inclined by 45\grad\ and rotating around a 2.0~\Msun\
  point source; the external and internal dashed lines correspond to
  stellar masses of 2.5~\Msun\ and 1.5~\Msun\, respectively . The
  horizontal and vertical straight dashed lines mark the systemic
  velocity  (5.8~km/sec) and the position of the continuum peak.}
\end{center}
\end{figure}

As shown in Fig.~\ref{fig:12CO},~\ref{fig:12CO32} and \ref{fig:13CO},
in all the detected molecular line transitions,
the emission is resolved, showing a velocity pattern typical of an
inclined rotating disk characterized by a position angle of about
130\grad\ (a more precise estimate of the position angle will be
presented in Sec.~\ref{sec:res}). The velocity-position diagram 
calculated along this direction (see Fig. \ref{fig:kep_rot}) shows a
well  defined ``butterfly shape'' typical of  Keplerian rotation. A
first estimate of the mass of the central 
object and of the dimension of   the disk can be obtained by comparing
the observed velocities with the Keplerian law:
\begin{equation}
\label{eq:kep}
v_{\epsilon} = C \cdot \epsilon^{-1/2}, 
\end{equation}
where $\epsilon$ is the angular distance from the central
star and $v_{\epsilon}$ is the component of the disk rotational
velocity along the line of sight. If the stellar mass \Mstar\ is
in solar units, the stellar distance $d$ in parsec and $\theta$ is the
disk inclination ($\theta=0$ means pole-on disk), the constant $C
\simeq 30 \sqrt{M_{\star}/d}\sin{\theta}$  is the component along the 
line of sight of the disk rotational velocity (in km/sec) at 
$\epsilon=1\arcsec$. As shown in Fig. \ref{fig:kep_rot} the  envelopes
of both the $^{12}$CO and $^{13}$CO emissions are in agreement with
$C \simeq 2.7\pm0.4$ km/sec. This value corresponds to a  stellar mass
of $2.0 \pm 0.5$ \Msun\, for an inclination of 45\grad\ and
d$=122$~pc. Note finally that the disk rotation is clearly observed at  
least up to a distance of about $4\arcsec$, corresponding to a minimum
disk outer radius of about 500~AU. We will discuss in more detail the 
determination of the stellar mass and the disk outer radius in
Sec. \ref{sec:res}, using a detailed model for the continuum and CO 
molecular emission.     

\subsection{Free-free contribution and spectral index}
\label{sec:ff}

\begin{figure}
\begin{center}
\includegraphics[width=6.4cm, angle=270]{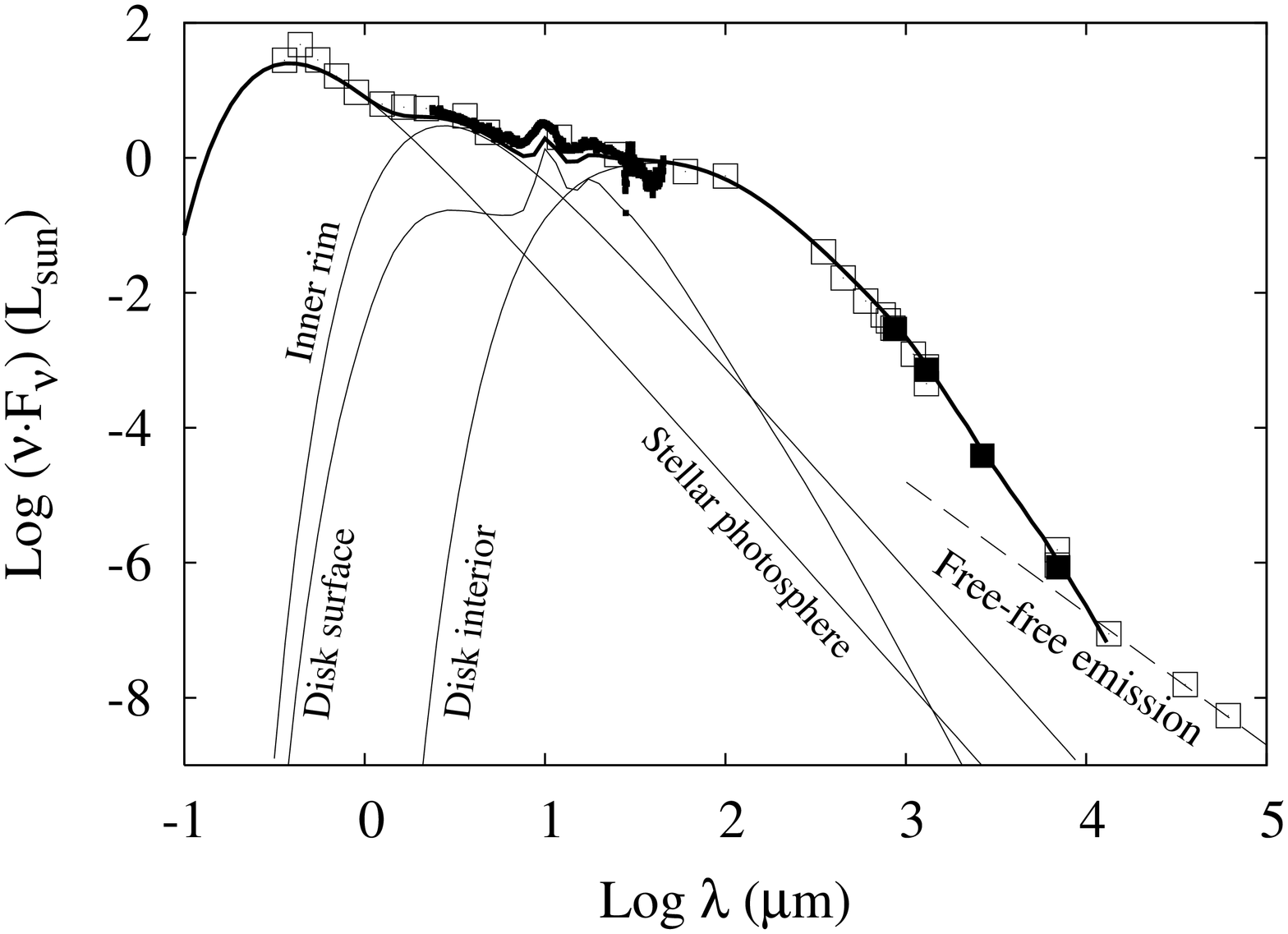} 
\caption{\label{fig:flux} HD~163296 spectral energy distribution. Data
  from literature are shown with empty squares (Mannings~\cite{Man94};
  Natta et al.~\cite{Nea04}; de Winter et al.~\cite{dW01}) and dots
  (ISO-SW). Our new measurements (see Tab.\ref{tab:obs}) are shown
  with full squares. The observed fluxes at 3.6~cm and 6.1~cm have
  been used to calculate the free-free contribution (dashed line) to
  the observed flux assuming a free-free spectral index of
  0.6. Free-free subtracted fluxes are shown at 7~mm 
  while at shorter wavelength the free-free contribution is
  negligible. The solid line shows the prediction of our disk model as
  discussed in Sec.~\ref{sec:res_cont}. The resulting spectral index 
  $\alpha_{mm}$ between 0.87 and 7~mm ($F_{\lambda} \propto
  \lambda^{-\alpha_{mm}}$) is $3.0 \pm 0.1$.}  
\end{center}
\end{figure}

The continuum spatially integrated fluxes of HD~163296 and the
corresponding spectral energy distribution are given in
Tab.~\ref{tab:obs} and shown in Fig.~\ref{fig:flux}, respectively. In
addition to our measurements (full squares), observations at 0.7, 1.3, 3.6  
and 6.1~mm are from Natta et al. (\cite{Nea04} and references therein)
while observations at 0.75, 0.8, 0.85, 1.1 and 1.3~mm are from
Mannings (\cite{Man94}).    

Assuming that the observed flux at 3.6 and 6.1~cm is dominated by  
free-free emission from a wind (with a spectral index of 0.6), the
free-free contribution at 7~mm  corresponds to 1.2~mJy (27\% of the
observed flux), while it is negligible at shorter wavelengths.  

After the subtraction of the free-free component, the millimeter
spectral index $\alpha_{mm}$ ($F_\lambda \propto
\lambda^{-\alpha_{mm}}$) calculated between 1~mm and 7~mm is $3.0 \pm  
0.1$, slightly higher than the value $\alpha_{mm} = 2.6 \pm 0.2$  
obtained by Natta et al.~(\cite{Nea04}) using VLA and OVRO fluxes
only.

\subsection{Disk mass}
\label{sec:obs_mass}

Assuming that the dust emission is optically thin at millimeter
wavelengths and that the dust is isothermal, the measured flux can be
used to estimate the product of the disk total mass $M$ times the dust
opacity  $k_{\nu}$, through the relation: 
\begin{equation}
\label{eq:Mdisk}
M \cdot k_{\nu} = \frac{F_{\nu}d^2}{B_{\nu}(T)}, 
\end{equation}        
where $d$ is the source distance and $T$ is the temperature of 
the emitting dust. At millimeter wavelengths, the dust opacity can by
parametrized by a power law
\begin{equation}
\label{eq:k_dust}
k =0.01 \cdot (\lambda/1.3\textrm{mm} )^{-\beta} \textrm{cm$^2$ g$^{-1}$},
\end{equation}
where the normalization at $\lambda=1.3$~mm
assumes a  dust/gas mass ratio of 0.01 (Beckwith et
al.,~\cite{BSC90},~\cite{BS91}). Taking a characteristic
temperature for the dust in the outer disk of an early A star of 30~K
(Natta et al.~\cite{Nea_PPIV}),we derive a total gas+dust mass of
0.12~\Msun\ for the circumstellar material in HD~163296. This rough
mass determination, that does not take into account the presence of 
optically thick emission from the inner part of the disk, nor of
temperature gradients, will be discussed in more detail in
Sec.~\ref{sec:res_cont}.    

With these simplifying assumptions, the slope of the dust
opacity is $\beta = \alpha - 2 \sim 1$, where $\alpha$ is the spectral
index obtained in the previous section.

\section{Disk Models} 
\label{sec:model}

The disk parameters derived in Sec.~\ref{sec:obs_res} under a number  
of very crude assumptions can only provide  order-of-magnitude
estimates. A more quantitative analysis requires to compare the
observations to  more sophisticated model predictions. We chose to
perform the comparison using the observed visibilities (rather than
reconstructed images), following a similar approach as the one
discussed by Dutrey et al.~(\cite{Dea_PPV}).

This method requires to decide ``a priori" which family of models  
is likely to describe the observed object. In view of the results
described in Sec. \ref{sec:obs_res}, we model  the millimeter
emission of HD~163296 (continuum and CO lines) as coming from a
circumstellar disk. We assume that the disk is heated by the stellar
radiation only, and that any viscous contribution can be
neglected. This is very likely a good approximation, given the
relatively low accretion rate measured in HD~163296 ($\sim 10^{-7}$
\Msun/yr; Garcia Lopez et al.,~\cite{GL06}). 

The disk structure and emission is computed using the 2-layer
approximation of Chiang and Goldreich (\cite{CG97}), as developed by
Dullemond et al.~(\cite{DDN01}). Similar models have been used in
Testi et al.~(\cite{Tea03}) and Natta et al.~(\cite{Nea04}), to
analyze the  (sub)millimeter emission of a number of Herbig Ae
stars. We refer to these papers for a more detailed description.

The inner disk is characterized by the presence of a {\it puffed-up}
inner rim located at the dust evaporation radius (Isella and Natta,
\cite{IN05}). The combined rim+disk flux is calculated taking into
account the shadow that the rim casts over the outer disk. 

The disk geometry can be fully flared, as in hydrostatic equilibrium
disks where gas and dust are well mixed. If dust growth and settling
are important, the flaring angle decreases: the resulting radial
temperature profile is flatter, affecting the SED in the
mid and far infrared (D'Alessio et al. \cite{DA06} and references
therein). As shown in Fig.~\ref{fig:flux}, the SED of HD~163296 is
well reproduced by a fully flared disk model. The fit of the SED
becomes very poor if the flaring angle is reduced by more than a
factor of two, which corresponds to a small variation on temperature 
radial profile. 

Once the stellar properties are known, the disk structure is
completely characterized by the following parameters: the disk mass
($M_d$), the disk outer radius ($R_{out}$), the
dependence of the surface density on  radius ($\Sigma \propto r^{-p}$)
and the properties of dust on the disk surface and midplane. In
addition, the observed emission depends on the orientation of the disk
with respect to the observer, which is characterized by  the
inclination $\iota$ of the disk with respect to the line of sight
($\iota=0$ for face-on disks) and the position angle $PA$.

\subsection{Continuum emission}
\label{sec:mod_cont}

The continuum emission at millimeter and sub-millimeter wavelengths
is computed  by ray integration as in Dullemond et al.~(\cite{DDN01}).    
We describe the midplane dust opacity  at long wavelengths as a
power law of index $\beta$, with $\beta$ a free parameter, as in
Eq. \ref{eq:k_dust}. At short wavelengths and in the disk surface we
adopt the dust opacity of astronomical silicates (Weingartner and
Draine, \cite{WD01}). The inner disk radius is the dust sublimation
radius, as in the rim models of Isella et al. (\cite{ITN06}) for 
large ($\sim 1 \mu$m) grains, and it is equal to 0.45~AU; the dust on
the disk surface is as in Natta et al.~(\cite{Nea04}). Neither of
these two quantities is relevant for the following analysis. 

\begin{table*}[!t]
\begin{center}
\begin{tabular}{lllll}
Parameters & Continuum &  $^{12}$CO J=2-1 & $^{12}$CO J=3-2 & $^{13}$CO
J=1-0 \\
\hline
\hline
\\ 
\vspace{0.1cm} $M_{\star}$ (\Msun) & 2.6$^a$ & $2.6^{+0.3}_{-0.5}$ &
2.4 $\pm$ 0.8 & 2.6 $\pm$ 0.6 \\ 
\vspace{0.1cm} $PA $ & 120\grad $\pm$ 20\grad &  128\grad $\pm$ 5\grad
& 130\grad $\pm$ 13\grad &130\grad $\pm$ 8\grad \\ 
\vspace{0.1cm} $Incl$ & 40\grad $\pm$ 12\grad  &  45\grad $\pm$ 5\grad
& 45\grad $\pm$ 10\grad & 50\grad $\pm$ 8\grad \\  
\vspace{0.1cm} $R_{out}$ (AU) & 200 $\pm$ 15 &  550 $\pm$ 50 &
550 $\pm$ 100 & 500 $\pm$ 80 \\  
\vspace{0.1cm} $\Sigma_{10AU}$ (g/cm$^2$) & 46 $\pm$ 4 &  90 $\pm$ 70
& 90$^a$ & 4$^{+12}_{-3}$ \\ 
\vspace{0.1cm} $p$ & 0.8 $\pm$ 0.1 & $0.6^{+0.3}_{-0.1}$ & 0.6$^a$ &
1.0 $\pm$ 0.5 \\ 
\vspace{0.1cm} $T_{CO,100AU}$ (K) &   & 40$^{+2}_{-5}$ & 60 $\pm$
20 & 30$\pm$ 10 \\
\vspace{0.1cm} $q$ & &  $0.5^{+0.2}_{-0.1}$ & 0.8$\pm$ 0.4 & 0.8$\pm$ 0.4  \\
\vspace{0.1cm} $\beta$ & 1.0  $\pm$ 0.1 & 1.0$^a$ & 1.0$^a$ & 1.0$^a$ \\
\vspace{0.1cm} $\chi^2_r $ & 2.6 &  1.17 & 1.05 & 1.15 \\

\hline
\end{tabular}
\caption{ \label{tab:res} Parameters of the disk structure relative to
  the best fit models for the continuum and the CO emissions as
  described in Sec.~\ref{sec:res}. For each parameters uncertainties
  are given at 1$\sigma$ level. $^{a}$ Fixed parameter.}   
\end{center}
\end{table*}

\subsection{CO emission}       
\label{sec:mod_CO}
The observed CO emission  originates in the outer layers of the disk,
at heights that depend on the optical depth of the specific
transition. Once the disk structure is specified, as described above,  
one needs to compute, at each radius, the gas  temperature  profile in
the vertical direction. This is a complex problem, whose results
depend on a number of not well known properties, among them the X-ray
field and the role of very small grains in heating the gas (e.g.,
Dullemond et al.~\cite{Dul_PPV}). Therefore, we use a parametric
description assuming that for each CO transition the excitation
temperature is the same at all $z$ and  can be described as a
power-law of $r$ in the form: 
\begin{equation}
T_{line} = T_{line}(r_0) (r/r_0)^{-q}.
\end{equation}
The assumption of a constant excitation temperature along the vertical  
direction is correct for optically thick lines, and/or if the
velocity gradient along the line of sight is very large. In both
cases, the contribution at each wavelength from the line of sight
under consideration comes from a small region only, where the optical
depth in the line is of order unity. For any CO line, the height to
which $T_{line}(r)$ refers is therefore different, depending on the
density and velocity structure, as well as on the disk inclination
angle. The values of $T_{line}(r_0)$ and $q$ for each CO transition
are free parameters. A similar procedure has been used by Dutrey
et al.~(\cite{Dea94}) and Dartois et al.~(\cite{Dea03}) in their
study of the CO emission of T Tauri disks. 

In the analysis of CO observations we have assumed $^{12}$CO/H$_{2}$ =
$7.0\cdot10^{-5}$, $^{13}$CO/H$_{2}$ = 1.0$\cdot10^{-6}$ and
C$^{18}$O/H$_2$ =1.3$\cdot10^{-7}$, which correspond to an isotopic
ratio $^{12}$CO/$^{13}$CO~=~70, $^{12}$CO/C$^{18}$O~=~550 (Beckwith
and Sargent, \cite{BS93}; Dutrey et al., \cite{Dea96}; and reference 
therein)  

As described in detail in the Appendix, we have developed a code which 
computes, for each CO transition, the line intensity and profile as
function of the disk parameters, namely the inclination and $PA$, the
density and temperature profile, the disk outer radius. In addition,
we assume that the disk is in Keplerian rotation around the central
star and vary the stellar mass independently for each line. 

In the following, we assume a gas turbulence velocity 
$v_{turb}=0$ km/sec. Note that for $v_{turb}$ in the 
range observed in other TTS and HAe (0.07-0.38 km/sec from
Pi\'etu et al.~\cite{Pea07},~\cite{Pea05},~\cite{Pea03}; Dutrey et
al.~\cite{Dea94},~\cite{Dea98}; Dartois et al.~\cite{Dea03}; Simon et
al.~\cite{Sea01}) the results does not change. This is due
to the fact that, for the inclination of the HD~163296 disk, the CO 
lines width is dominated by the differential disk rotation.

\section {Results}
\label{sec:res} 

\subsection{Method of analysis}
\label{sec:res_ana}

The observations of HD~163296 have been analyzed by comparing the
observed and the model predicted complex visibilities. For each set of
data, the best fit model has been obtained minimizing the $\chi^2 = 
\Sigma_i((Re_{mod,i}-Re_{obs,i})^2+(Im_{mod,i}-Im_{obs,i})^2)\cdot
W_i$, where $Re_X$ and $Im_X$ are the real and the imaginary part of
the complex visibility measured ($obs$) and predicted ($mod$) in the
point $i$ of the uv-plane, and $W_i$ is the weight of each
measure. For each CO transition, the $\chi^2$ has been computed by the
simultaneous fitting of 14 velocity channels chosen in order to
optimize the sampling of the line profile. The $\chi^2$  minimization
has been performed exploring a wide region of the space of model
parameters. For each parameter, the 1$\sigma$ uncertainties  are
estimated as $\chi^2_{1\sigma}= \chi^2_m + \sqrt{2n}$, where $n$ is
the number of degrees of freedom and $\chi^2_m$ is the $\chi^2$ value
of best fit model.

\subsection{Continuum emission}
 \label{sec:res_cont}

\begin{figure*} 
\begin{center} 
\includegraphics[width=5cm, angle=270]{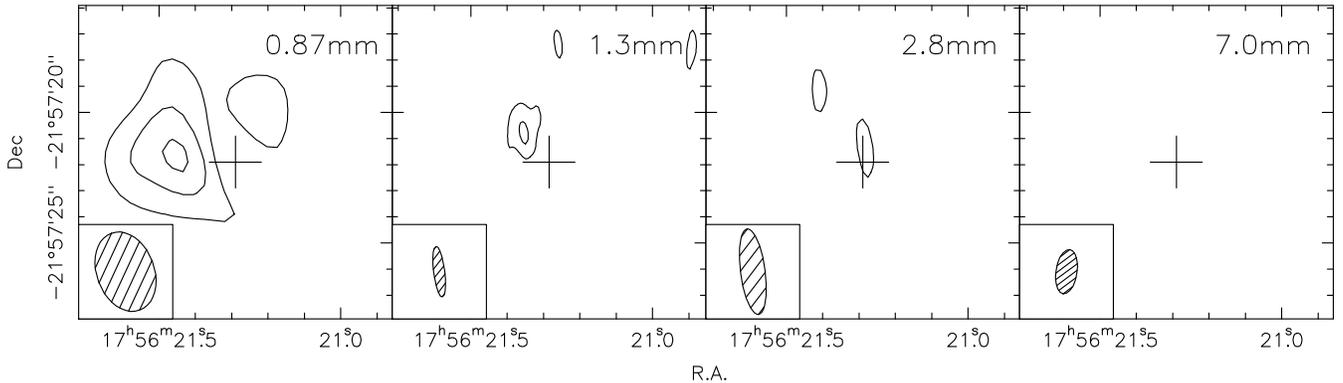} 
\caption{ \label{fig:res_cont}  Maps of the residuals relative to the
  best fit model for the continuum emission (see Fig. \ref{fig:cont}
  for the observations). The contour level are spaced by $3\sigma$
  corresponding to 30~mJy at 0.87~mm, 12~mJy at 1.3~mm, 3.3~mJy at
  2.8~mm and 0.75~mJy at 7~mm. The small boxes show the relative
  synthesized beams. } 
\end{center}
\end{figure*}

The continuum data at the four observed wavelengths
(0.87,~1.3,~2.7,~7~mm) have been analyzed independently to estimate
the best disk parameters and their uncertainties. We have varied the two 
parameters which define the position of the disk on the plane of the
sky, namely the inclination and position angle, as well as the three
physical  parameters which affect the continuum dust emission, namely
the disk outer radius $R_{out}$, the slope $p$ of the surface density
profile and the product $\Sigma_{10AU} \times k_\lambda$. 
Since the continuum emission has a very weak dependence on the mass of
the central star, we fixed \Mstar=2.6 \Msun\ (see
Sec.~\ref{sec:res_CO}).  

The values of the parameters obtained at different wavelengths are the
same within the uncertainties. In column 1 of Tab ~\ref{tab:res} we
show the values of the resulting best fit model obtained by combining
the independent results at the four different wavelengths. The
parameters constraint is dominated by the 1.3~mm data, which have the
best sensitivity and resolution. 

The surface density radial profile has a slope $p = 0.81 \pm 0.01$ and
the  disk outer radius is $R_{out}=200\pm15$~AU. The very good 
constraints on both  parameters is due to the favorable orientation
of the beam at 1.3 mm with the maximum resolution (0.42$\arcsec$) in 
almost the same direction of the  major axis of the disk. 
Note that the formal error on $p$ is extremely low. On the other
hand, small variations on the disk flaring (that allow to fit the SED)
lead to correspondingly small variation of the disk radial temperature
profile and on the value of $p$. Taking into account this fact, $\pm
0.1$ is a more reasonable uncertainty for $p$. The formal
uncertainties on the other parameters are much larger than those
introduced by small variations of the flaring angle.

The four values of $\Sigma_{10AU} \times k_\lambda$ are used to
constrain the slope $\beta$ of the dust opacity law (see
Eq.~\ref{eq:k_dust}) between 0.87 and 7~mm, which turns out to be  $1.0 \pm
0.1$.  This  confirms the presence of large grains in the HD~163296
circumstellar disk (Natta et 
al.~\cite{Nea04}). Note that the optically thick disk region at 1.3~mm
has size of $\sim6$~AU and contributes only 4\% of the observed 
flux, so that $\beta$ is very similar to that derived in
Sec.~\ref{sec:ff}. The observations constrain only the dependence of
the dust opacity on wavelength, not its absolute value. The value of
$\Sigma_{10AU}$ given in Tab.~\ref{tab:res} is obtained from 
$\Sigma_{10AU} \times k_{1.3\textrm{mm}}$ assuming
$k_{1.3\textrm{mm}}=0.01$ cm$^2$/g.  

The residuals for each wavelength
are shown in Fig.~\ref{fig:res_cont}. They have been reconstructed
from the residuals in the uv-plane with the same procedure used to
obtain the images in Fig.~\ref{fig:cont}. Residual contours are
generally lower than $3\sigma$ with the exception of the 0.87 and
1.3~mm maps, where a flux asymmetry in the east half of the map (see
also Sec.~\ref{sec:obs_morf}) is clearly visible. This structure, not  
detected at longer wavelengths, requires more resolved observations to 
be investigated in detail.

\subsection{CO emission}
\label{sec:res_CO}

The analysis of the CO emission has been carried out separately for
the different CO transitions and the corresponding best fit parameters
are given in Tab.~\ref{tab:res}. 

\begin{figure}[!t] 
\begin{center} 
\includegraphics[width=8cm, angle=0]{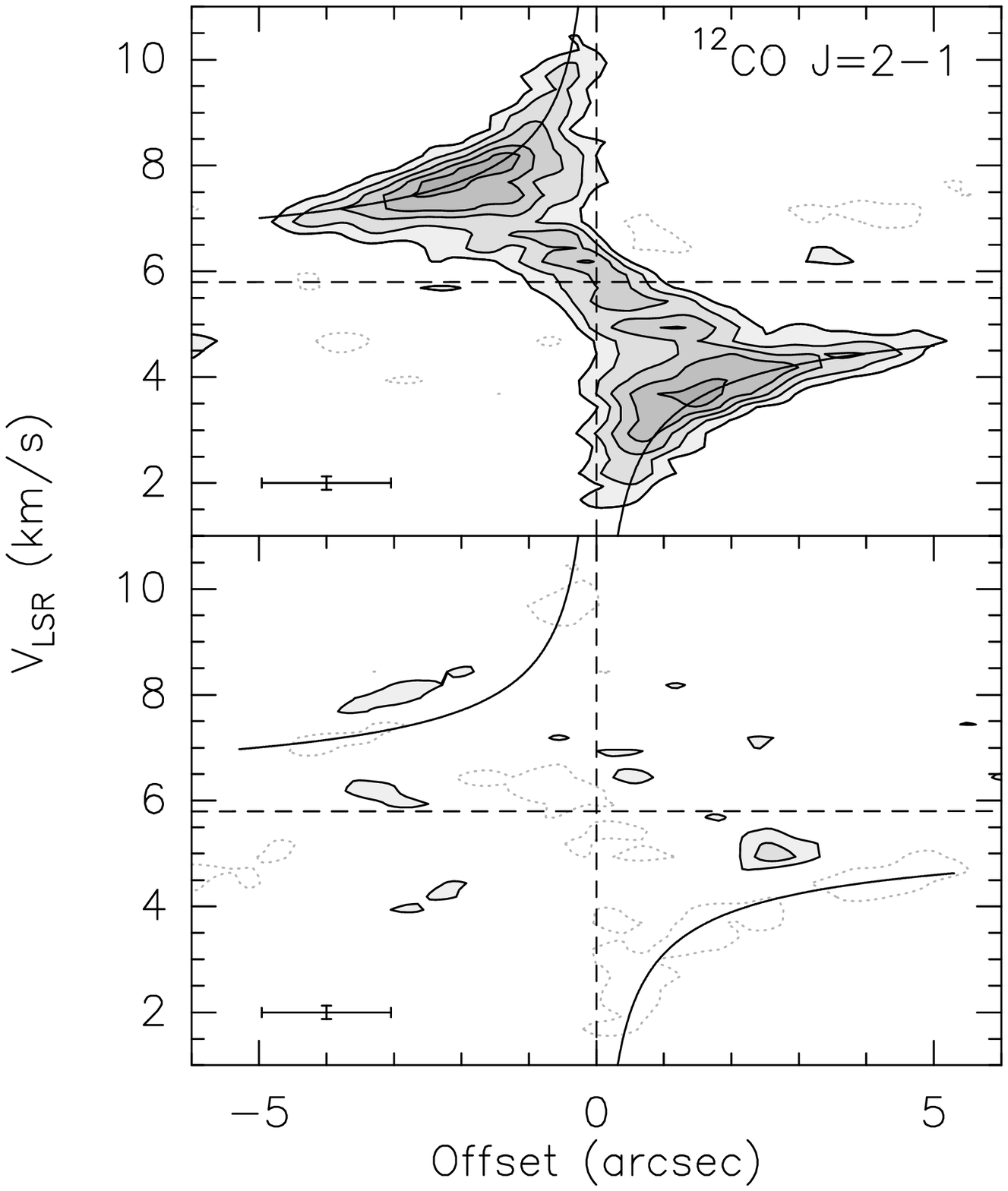} 
\caption{ \label{fig:res_velo} Comparison between the observed and the 
  model predicted $^{12}$CO J=2--1 emission. The upper panel shows the 
  position-velocity diagram for the $^{12}$CO J=2--1 transition (as in
  Fig.~\ref{fig:kep_rot}). The lower panel shows the residuals relative
  to the best fit model parameters reported in Tab.~\ref{tab:res}. The
  contour levels are at 2$\sigma$ as in Fig.~\ref{fig:kep_rot}.
  }
\end{center}
\end{figure}

The observed $^{12}$CO J=2-1 emission is well fitted by a Keplerian
disk orbiting a central star with mass
$M_{\star}=2.6^{+0.3}_{-0.5}$~\Msun. Fig.~\ref{fig:res_velo} shows the     
position-velocity residuals obtained subtracting the best fit model
from the observed uv--table: no evidence of non-Keplerian rotation or
stellar outflow is detected, within the actual instrumental
sensitivity. Both the position angle and the
inclination are in agreement with the values
obtained from the continuum.  Since the line is optically thick, the
constraint on the gas surface density is poor. The 
inferred outer radius of the disk is $550 \pm 50$~AU, more than two
times larger than the value obtained from the continuum and similar to
the result obtained by Thi et al. (\cite{TZD04}) from the model
fitting of single dish line profiles. The
radial temperature profile of the $^{12}$CO has a slope
$q=0.5^{+0.2}_{-0.1}$ and a value of $40^{+2}_{-5}$~K at 100~AU.    

Since the $^{12}$CO J=3--2 line is optically thick and given the
relatively low resolution of the SMA observations, the
measurements do not constrain the CO radial density profile. In this
case we fix both $p$ and $\Sigma_{10AU}$ equal to the values obtained
for the $^{12}$CO J=2--1 and vary the other parameters. The 
results for the inclination, position angle, disk outer radius and CO 
temperature are in good agreement with the values obtained for the
$^{12}$CO J=2--1, with larger uncertainties due to the lower spatial
resolution. 

Finally, the model fit to the $^{13}$CO J=1--0 line gives results
consistent with those obtained from the $^{12}$CO lines with the exception
of the value of $\Sigma_{10AU}$ which is significantly smaller. This 
discrepancy may be due to a depletion of the $^{13}$CO and will be
discussed in Sec.~\ref{sec:disk_CO}. The radial temperature profile of
the $^{13}$CO has a slope $q=0.8 \pm  0.4$ and a value of $30\pm 10$~K
at 100~AU.  We will comment on the gas physical conditions in
Sec.~\ref{sec:disk_CO}. In general, the parameter constraints obtained 
from the $^{13}$CO are not so good as those obtained for the $^{12}$CO
J=2--1, as expected given the lower resolution of the observations.

\section {Discussion}
\label{sec:disc}

\subsection{Disk orientation}
The disk orientation is well determined by our set of
measurements: it has it has a moderate inclination with respect to the
line of sight (the mean of the inclination values is
46\grad~$\pm$~4\grad) with a position angle of 128\grad~$\pm$~4\grad. 


\subsection{Disk kinematics and stellar mass}
All the observations are consistent with the emission of a
circumstellar disk   in Keplerian rotation around a star of
2.6~\Msun, assuming the Hipparcos distance of
$122^{+17}_{-13}$~pc (van den Ancker et al.~\cite{vdA98}). Within the
error, the stellar mass is in agreement with the value of 2.3 \Msun\
(Natta et al.,~\cite{Nea04}) obtained from the location of the star on
the HR diagram, using Palla and Stahler (\cite{PS93}) evolutionary
tracks; the corresponding stellar age is of about 5~Myr.

\subsection{Disk outer radius}
\label{sec:out_rad}

\begin{figure}[t]
\includegraphics[angle=0 ,width=9cm]{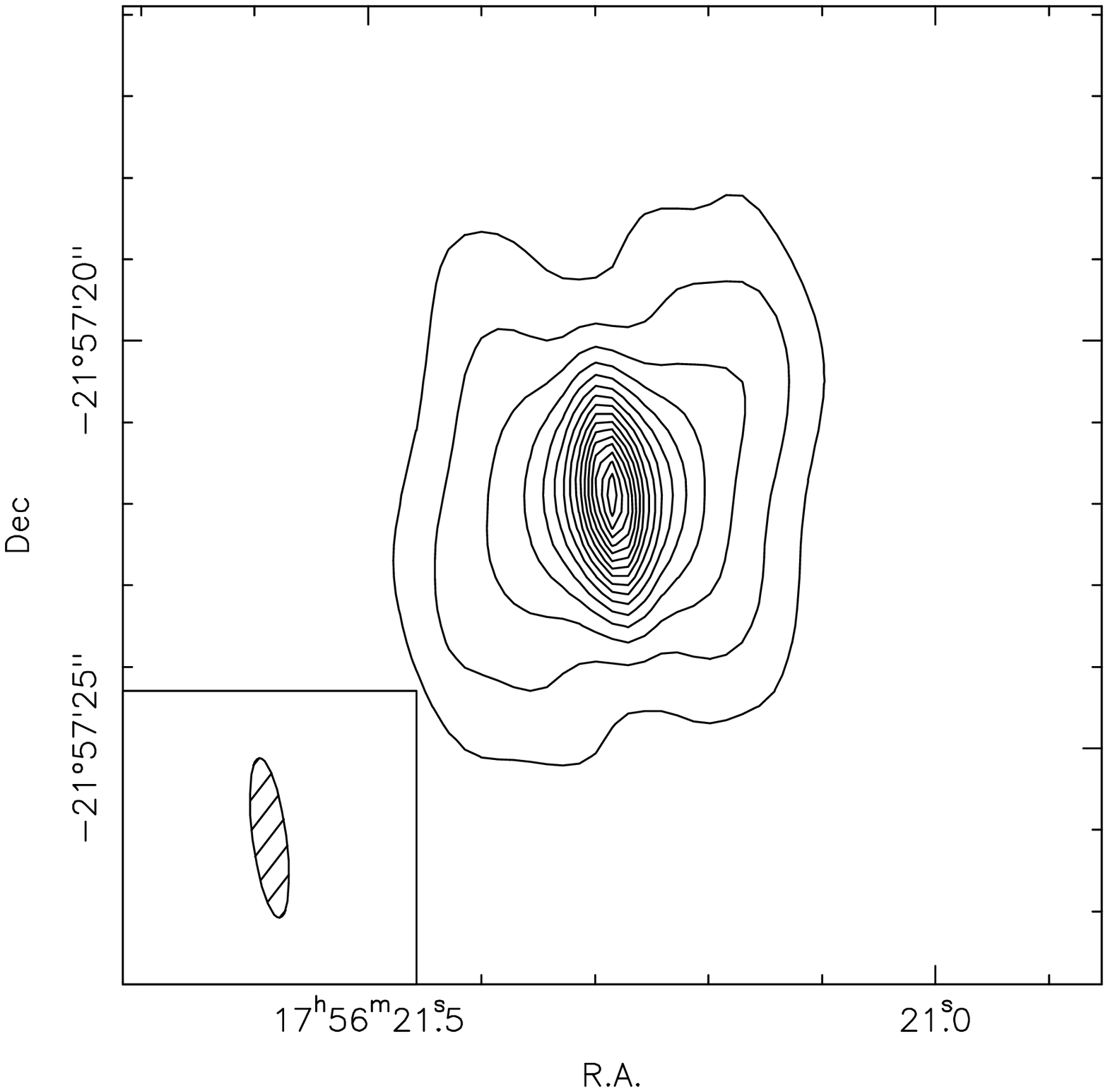} 
\caption{\label{fig:cont_550} Model predicted 1.3~mm continuum
  emission obtained extrapolating the dust surface density inferred
  inside the radius of 200~AU to the CO outer radius of 550~AU. The
  contours are spaced by 3$\sigma$ corresponding to 0.12~mJy. See
  Fig.~\ref{fig:cont} for the comparison with the observations.
}
\end{figure}


The model fitting (Tab.~\ref{tab:res}) shows that the value of the disk 
outer radius inferred from the continuum dust emission
(200~$\pm$~15~AU) is almost three times smaller than the value obtained 
from the CO analysis (where 540~$\pm$~40 AU is the mean value). Since
our method takes into account the sensitivity limits of the different
interferometric observations, this discrepancy can not be explained by
the fact that the disk outer regions have a continuum surface
brightness below the sensitivity limit. If we extend
the disk model that fit the continuum to the outer radius of the CO,
we predict a continuum emission between 200~AU and 550~AU which would
be easily detected in our observations (see Fig.~\ref{fig:cont_550}). A
similar difference in the dust and gas outer radii has been also
reported by Pi\'etu et al.~(\cite{Pea05}) for the Herbig Ae star AB Aur.  


To reconcile CO and dust observations, it is necessary to
introduce a sharp drop in the continuum emission of a factor $>
30$ at a radius of about 200~AU. With such a drop, the millimeter
fluxes at  larger $r$ will be below the sensitivity of our
observations and will be lost in the observational noise. What can be
the origin of such a drop?   

In the optically thin regime, which is appropriate for the outer
regions of the disk, the continuum flux emitted at distance
$r$ from the star depends on the mass surface density $\Sigma(r)$, the
dust/gas ratio $\Pi(r)$, the dust opacity $k_\nu(r)$ and the midplane
dust temperature $T(r)$ through the relation
\begin{equation}
\label{eq:flux_dep}
F_\nu(r) \propto \Sigma(r) \cdot \Pi(r) \cdot k_\nu(r) \cdot T(r).
\end{equation}

It is possible that a different disk geometry (e.g., a lower flaring
angle) gives a midplane dust temperature lower than the values
predicted by our disk model. A lower temperature limit, of $\sim$10~K,
is however imposed by the equilibrium with the interstellar radiation
field. Given that our fully flared disk model predicts temperatures of
20-30~K in the millimeter emitting regions, one can reduce $F_{\nu}$
by a factor 3 at most. 

A second possibility to explain the observed flux depletion is that
the dust opacity $k_\nu (r)$ at distance larger than 200~AU is much
lower because most of the grains have grown into very large
bodies. 

Finally, a third possibility is that the dust density, i.e. $\Sigma(r) 
\cdot \Pi(r)$, is in some way depleted at large distances. The values
of the gas density $\Sigma(r)$ obtained by the CO 
lines analysis (see Tab.~\ref{tab:res}) induce however to exclude the
existence of the strong discontinuity in the gas radial density
profile required to explain a flux depletion factor $> 30$ since a
drop of a factor $\sim$10 in the $^{13}$CO 
distribution at $r>200$~AU will already produce small but observable
effects on the line emission. On the other hand, it is possible that
the ratio dust/gas $\Pi(r)$ decreases very 
rapidly in the outer disk either due to the formation and fast
migration of meter size bodies or to the presence of planetesimals or
planets which create a gap in the dust distribution. It is worth to
note that the dynamical perturbation induced by such large bodies on
the surrounding material (see the review  of Papaloizou et
al.,\cite{PP07}) may also account for the asymmetric dust density
distribution detected in the continuum maps
(Fig.~\ref{fig:res_cont}).  

The presence of a giant planet, or a brown
dwarf, orbiting in the outer disk of HD~163296 has also be suggested
by Grady et al.~(\cite{G2000}), in order to explain the dark line
observed in the scattered light HST images between 300~AU and 350~AU
from the central star. While planets are invoked to explain the large
inner gaps observed in the dust distribution in the so called
``transitional disks'' (i.e., Calvet et al.,~\cite{C05}), HD~163296
will be, if confirmed by future observations, the first case in which
a sub-stellar mass companion is found in the outer disk of a pre-main
sequence star.

\subsection{Disk surface density and mass}
Both the dust continuum emission and the CO lines indicate a rather  
shallow surface density profile, $\Sigma \propto r^{-0.8}$
corresponding to a gas~+~dust mass of 0.05~$\pm$~0.01~\Msun\  if the
disk outer radius is 200~AU. However, if, as we believe, the gaseous
disk extends with the same surface density profile all way to 540~AU
the inferred disk mass is much larger, 0.17~\Msun. 

An additional uncertainty on the disk mass comes from the
uncertainty on the dust opacity, as discussed in Natta et
al.~(\cite{Nea_PPV}). Although a disk mass of 0.17~\Msun (6\% \Mstar)
is probably a lower limit, the observed Keplerian rotation pattern
indicates that the disk should not be much more massive. In this case
one expects deviations from Keplerian rotation (Lodato and Bertin,
\cite{LB03}) as observed in massive disks (Cesaroni et al.,
\cite{C_PPV}).  

The example of HD~163296 illustrates how the simple disk mass
derivation from the observed millimeter flux (see
Sec~\ref{sec:obs_mass}) can be wrong for a variety of reasons, even 
when the emission is optically thin. In the case of HD~163296, only
30\% of the total mass (i.e, the fraction inside 200 AU)
contribute to the continuum millimeter flux. On the other hand, the
disk is hotter than the 30~K assumed in Eq.~\ref{eq:Mdisk}. The two
effects cancel in part so that the ``simple'' and ``correct'' values
differ by about 50\% (for the same $k_{1.3\textrm{mm}}$
normalization).

\subsection{CO vs dust temperature}
\label{sec:disk_CO}

\begin{figure}[t]
\includegraphics[angle=270, width=9.3cm]{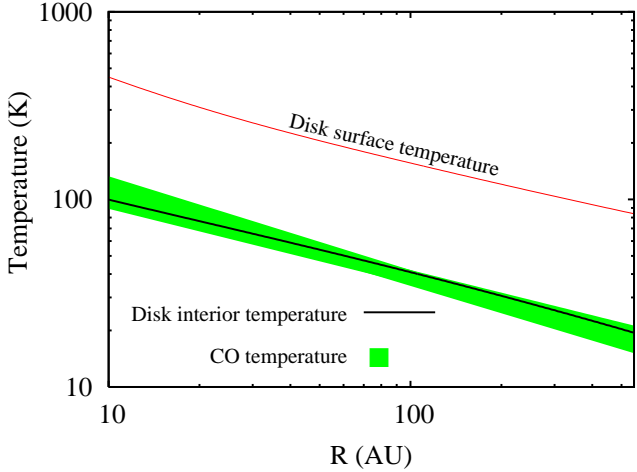}
\caption{\label{fig:COtemp} Comparison between the dust and the CO
  temperature. As labelled in the figure, the thin-solid line
  correspond to the dust temperature in the optically thin disk
  surface, while the thick solid line corresponds to the dust
  temperature in the disk interior. The CO
  temperature is indicated with the shaded region, taking into
  consideration the uncertainties. The model parameters are those
  reported in column~3 of Tab.~\ref{tab:res}}
\end{figure}

At the actual gas density both the $^{12}$CO and $^{13}$CO lines
are optically thick and lead to similar CO excitation temperatures
expressed by the relation $T(r) \sim 40$K$\cdot (r/100$AU$)^{-0.5}$.  
Fig.~\ref{fig:COtemp} shows the comparison between this CO temperature  
(shaded region) and the dust temperature corresponding to the same
disk model. The disk interior temperature $T_i$ (thick-solid
line) and the disk surface temperature $T_s$ (thin-solid line),
calculated using our two-layer disk model, correspond respectively to
disk regions where $\tau_{\star} >> 3$ and $\tau_{\star}\simeq
1$ (Dullemond et al.~\cite{D02}); $\tau_{\star}$ is the optical depth
for the stellar radiation calculated along the radial direction. The
equivalence between the CO and the dust temperature can thus be
interpreted as the evidence that the CO, and in particular the
$^{12}$CO, lines are emitted by the gas present at such values of the
optical depth, e.g. well under the disk surface. Is this result
compatible with the high optical depth of the $^{12}$CO lines?    

Assuming that H$_2$ and CO are well mixed all over the vertical
extension of the disk and assuming also the standard ratio
$^{12}$CO/H$_2$=7.5$\cdot10^{-5}$, the $^{12}$CO lines should be
emitted by disk regions well above the disk surface, defined as the
height where $\tau_{\star}=1$. In this situation, we expect a CO 
temperature close to the disk surface temperature, or  
even higher if the gas and the dust are decoupled (Jonkheid et al., 
2006). On the other hand CO molecules on the superficial layers of the 
disk are expected to be photo-dissociated by the UV photons emitted by 
the hot central star. In particular, for an Herbig Ae star similar to
HD~163296, Jonhkeid et al. (2006) show that the CO 
photo-dissociation occurs in the disk regions where the optical depth
to  the stellar UV radiation field is $\tau_{UV} \simless 3$ (see Fig
5.6 therein). Therefore, the CO and disk interior temperature are 
expected to be similar if $\tau_{UV} \simless \tau_{\star}$, i.e.,
if $k_{UV} \simless k^P_{\star}$, where $k^P_{\star}$ is the dust
opacity averaged on the stellar spectrum. We think that this is the 
case in HD~163296. 

\begin{figure}[t]
\includegraphics[angle=270, width=9cm]{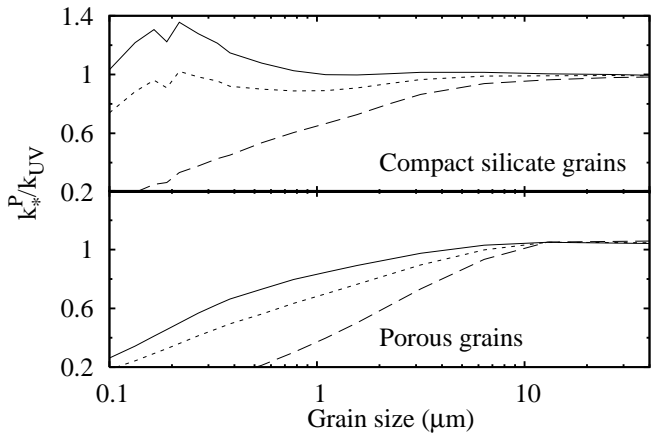}
\caption{\label{fig:kratio_av} 
  Ratio between the Planck mean opacity at
  the stellar temperature $k^P_{\star}$ and the UV opacity $k_{UV}$
  for different grain sizes and composition. The upper panel shows the
  opacity ratio for compact silicate grains; the lower panel shows
  the ratio for porous grains composed by Olivine (17\%), carbonates
  (53\%) and vacuum (30\%). The solid lines are relative to an HAe
  star of $T_{\star}=10000$K, the 
  long-dashed lines refer to a TTS with $T_{\star}=4000$ K and the
  short-dashed lines to the same TTS with the addition of the UV
  excess radiation calculated so that $L_{UV}/L_{\star}=0.05$. This
  latter case is representative for the emission  released by the gas
  accreting on the stellar surface. 
}
\end{figure}

The dust opacity is a complex quantity that depends on the dust
chemical composition, structure and on the grain size
distribution. The condition $k_{UV} \simless k^P_{\star}$ is generally
satisfied by large grains. In the case of HD~163296 one needs, for
example, compact silicate grains larger than 0.1$\mu$m or porous
grains of silicate and carbonaceous materials larger than few microns 
(see Fig.~\ref{fig:kratio_av}). Cooler stars require larger grains,
unless significant UV excess is present.

If CO survives only in the disk interior, we expect that lines with
different optical depth will have similar temperatures. On the
contrary, if CO is present in the disk surface layers, lines with
higher optical depth should have higher excitation temperatures.

The few existing observations indicate that all cases occur. 
Pi\'etu et al.~(2007) have recently reported that for the 
HAe stars MWC~480 and AB~Aur and the TTS DM Tau, the $^{12}$CO and
$^{13}$CO lines are characterized by different temperatures, while in
the case of the TTS LkCa~15 the CO temperatures are all similar. We
suggest that these differences are due to different dust properties
(i.e. different composition, grain growth and settling) and stellar
spectra. 

\subsection{CO isotopic abundances}

The values of $\Sigma_{10AU}$ reported in Tab.~\ref{tab:res} indicate
that the $^{13}$CO J=1--0 emission requires a ratio
$^{13}$CO/H$_2$~$\sim10^{-7}$, about a factor 10 lower than what found in
interstellar clouds. From the non-detection of the C$^{18}$O
J=1-0 line we estimate that the C$^{18}$O should be depleted of a
factor $> 60$ with respect to the typical ratio
$^{12}$CO/C$^{18}$O~=~550.  
We note that a similar trend of depletion from $^{12}$CO to $^{13}$CO
and C$^{18}$O has been found by Dutrey et al. (\cite{Dea94},
\cite{Dea96}) in a number of TTS. On the other hand 
an over-abundance of $^{13}$CO has been found in DM~Tau, MWC~480 and
LkCa~15 (Pi\'etu et al. 2007) while for
DM~Tau, Dartois et al. (2003) find that all the CO isotopomers
($^{12}$CO, $^{13}$CO and  C$^{18}$O) are depleted by the same factor
$\sim$10. The observational picture is still very uncertain and
deserves further investigations.

In the case of HD~163296, the gas temperature always
higher that 20~K roles out the condensation of CO onto dust
grains as possible cause of the CO isotopomers depletion. Alternatively,
Jonkheid et al. (\cite{JD06}) have recently pointed out that an higher
ratio between the $^{12}$CO and its isotopomers may be the consequence of
grain growth and settling. This latter hypothesis is more convincing
given the evidence of grain growth in the HD~163296 disk discussed in
this work (see Sec.~\ref{sec:res_cont} and \ref{sec:disk_CO}).

\section{Summary}
\label{sec:sum}
This paper presents new observations of the disk of HD~163296 in the   
dust continuum from 0.87 to 7~mm, $^{12}$CO (J=2-1 and J=3-2) and
$^{13}$CO (J=1-0) lines. The disk is resolved in all lines and
continuum.  

We have compared the observations to the predictions of
self-consistent disk models. We find that the disk, as seen in CO
lines, is very large ($R=540$ AU), with  a Keplerian rotation
pattern consistent with a central mass of 2.6~\Msun. Within the
observational errors, there is no evidence of non-Keplerian motions
and/or significant turbulent broadening. We obtain a disk inclination
of 46\grad, significantly lower than the value of $\sim$60\grad\ found
in literature.

The dust opacity has a power law dependence on wavelength $\kappa
\propto \lambda^{-\beta}$ with $\beta=1.0\pm 0.1$ in the interval
0.87-7 mm. This value is similar to what has been measured in a number
of spatially resolved disks of HAe and TTS (e.g., Natta et
al.~\cite{Nea04}, Rodmann et al.~\cite{Rod06}), and is very
likely an indication that the bulk of the solid material in these
disks has coagulated into very large bodies, of millimeter and
centimeter size (Natta et al.~\cite{Nea_PPV}). Within the accuracy
of our data we do not confirm the possible variation of $\beta$ with
$r$ discussed in Natta et al. (\cite{Nea_PPV}).  

The continuum observations constrain the surface density profile
($\Sigma \propto r^{0.8}$) for $r \leq 200$ AU. At larger radii,
the continuum emission drops with respect to the model predictions  by
a factor 30 at least (at 1.3 mm). We argue that this may be due to the
clearing of a very large gap by dynamical perturbations from a low
mass companion or to the inward migration of the large bodies that may
form in the outer disk. 

The temperatures derived for CO lines of different optical depth are
similar and 
equal to the dust temperature in the disk interior ($\tau_{\star} >>
1$). This requires that the dust
opacity  in the UV (which controls the CO dissociation) and in the
wavelength range where the stellar radiation peaks are similar, as
expected if grains have grown to micron size.
Differences in the temperature of the CO lines can only be expected if
CO is present in the disk surface layers. A variety of situation can
occur depending on the exact dust composition and sizes and on the
stellar radiation field. 

A comparison of the disk properties derived from the dust continuum  
and the CO lines shows that the $^{13}$CO J=1-0 emission is consistent
with a ratio of $^{13}$CO/H and C$^{18}$O/H about a factor 10 and 60
lower than what is found in interstellar clouds.

\section{Conclusions}
\label{sec:conc}
The results discussed in this work can be interpreted as clues of
the evolution occurring in the HD~163296 system. The presence of large
grains in the disk midplane, the equivalence between the dust and CO
temperature, the drop in the continuum dust emission further out
200~AU, its detected asymmetry and, maybe, the CO isotopomers
depletion, all support the idea that the circumstellar disk is
probably harboring the formation of large bodies, being in between a
Class II pre-main sequence disk and an older debris disk.

On the other hand, it is important to underline that HD~163296
strongly differs from the so called {\it transitional disks} (Calvet
et al.,~2005) characterized by a clearing of the inner disk supposed
to be originated by the presence of a giant planet. In the case of
HD~163296 is indeed the outer disk that appear dust depleted. This
conclusion opens a number of theoretical problem about how dust grains
can growth in the low density outer disk environment. In this respect
the recently improved PdB array, the new CARMA array and ALMA will
bring in the next future to  real observational breakthrough in the
comprehension of circumstellar disk evolution and planetary
formation.

\begin{acknowledgements}
The authors acknowledge partial support for this project by  MIUR PRIN
grant 2003/027003-001. A.I. acknowledge Riccardo Cesaroni, Malcolm
Walmsley, Giuseppe Bertin, Ewine van Dishoek and Anneila Sargent for
their help and useful suggestions.
\end{acknowledgements}

\begin{appendix}
\section{CO emission model}       
\label{app:mod_CO}

In order to analyze the CO emission we developed a numerical code
that solve the general formulation of the radiation transfer
equation along each direction between the observer and the
emitting source. If $s$ is the linear coordinate along the line of
sight, increasing from the observer ($s \equiv 0$) towards the source,
the observed emission  in each direction is given by the relation 
\begin{equation}
I_{\nu} = \int_{0}^{\infty} S_{\nu}(s) e^{-\tau_{\nu}}(s)
  K_{\nu}(s) ds,
\end{equation}  
where the optical depth is defined in each point $s$ through 
\begin{equation}
\tau_{\nu}(s) = \int_{0}^{s}K_{\nu}(s')ds'
\end{equation}
and $K_{\nu}$ is the absorbing coefficient of the interstellar
medium. Given the high gas densities in the pre-main sequence
circumstellar disks, we can assume that all the CO levels
corresponding to the rotational transitions under investigation are
thermalized. In this case the source function can be approximated by
the Planck function  
\begin{equation}
S_{\nu}(s) = B_{\nu}(T_{CO}(s)) =
\frac{2 \textrm{h} \nu^3}{ \textrm{c} ^2}\frac{1}{\exp({\textrm{h}\nu/
    \textrm{k} T_{CO}(s)})-1},  
\end{equation} 
depending only on the local temperature of the gas $T_{CO}$ (c, h and
k are respectively the light speed, the Planck and Boltzman constant).

The absorbing coefficient of the circumstellar medium is due both to
gaseous CO and dust: $K_{\nu}(s) = K_{\nu}^{CO}(s)+K_{\nu}^d(s)$. For
the dust $ K_{\nu}^d(s)=\rho(s) \cdot k_{\nu} $, where $\rho(s)$ is
the local density of the circumstellar material (gas+dust) and $k_\nu$
is the dust absorbing coefficient for total mass unit given in
Eq.~\ref{eq:k_dust}. The CO absorbing coefficient is given by the
relation 
\begin{equation} 
 K_{\nu}^{CO}(s) = n_l(s) \cdot \sigma_{\nu}(s)  
\end{equation}
where $n_l(s)$ is the total number of CO molecules at the lower
level $l$ of the transition and $\sigma_{\nu}(s)$ is the CO absorbing
cross section. 

\begin{figure}[!t]
\begin{center}
\includegraphics[width=9cm, angle=0]{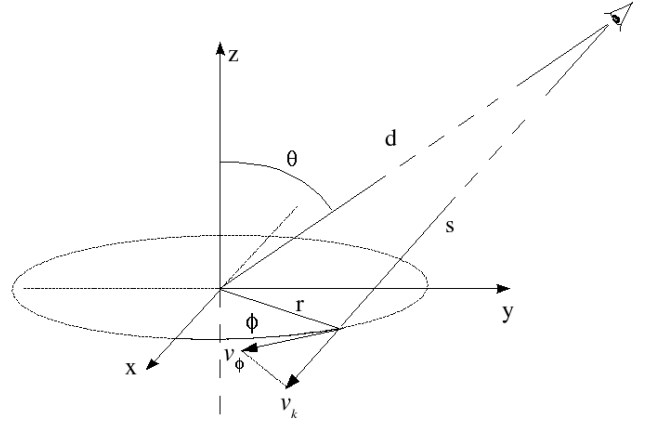} 
\caption{ \label{fig:disk} Schematic representation of the frame of
  reference adopted to calculate the CO emission arising from a
  rotating disk. The disk midplane and the observer lie respectively
  on the $(x,y)$ and $(y,z)$ planes; $\theta$ is the disk inclination;
  $d$ is the distance between the observer and the central star; $s$
  is the linear coordinate along the line of sight increasing from  
  the observer ($s \equiv 0$) towards the emitting source. Assuming
  that the material within the disk is subject to the Keplerian
  rotation around the central star, we call $v_{\phi}$ the velocity
  of a mass element at distance $r$ and $v_k$ the projection of
  $v_{\phi}$ along the line of sight.}
\end{center}
\end{figure}  

Calling $m_0$ and $\chi_{CO}$ the mean molecular weight of the gas and 
the fraction of CO present in the gas respectively, the number of
molecules $n_l(s)$ is given by the Boltzman equation 
\begin{equation}
n_l(s) = \chi_{CO} \frac{\rho(s)}{m_0}\cdot \frac{g_l \, e^{-E_l /
    \textrm{k}T_{CO}(s)}}{Z(T_{CO}(s))}, 
\end{equation}
where $g_l=2l+1$ is the statistical weight of the lower level $l$ of 
the transition, $E_l=(1/2)l(l+1)\textrm{k}T_1$ is the level energy,
$T_1$ is the temperature equivalent to the transition energy and
$Z(T_{CO}(s))$ is the partition function at the gas temperature
$T_{CO}(s)$.  Following Beckwith and Sargent (\cite{BS93}, and
references therein), the absorbing cross section $\sigma_{\nu}(s)$ can
be expressed in term  of the integrated cross section of the
transition $\sigma_0$ through the relation
\begin{equation}
\sigma_{\nu}(s) = \sigma_0 \cdot \phi_{\nu}(s) \cdot (1-e^{-\textrm{h} \nu
  / \textrm{k} T_{CO}(s)}),
\end{equation}
where $\phi_{\nu}(s)$ is the intrinsic line profile 
\begin{equation}
\label{eq:phi}
\phi_{\nu}(s) = \frac{\textrm{c}}{\nu_0} \cdot\, \frac{1}{\Delta V
  \sqrt{\pi}} \cdot\, 
\exp{\left(-\frac{\Delta^2v}{\Delta^2 V} \right)}, 
\end{equation}
with
\begin{equation}
\label{eq:vturb}
\Delta V = \sqrt{\frac{2\textrm{k}T_{CO}(s)}{m_{CO}}+v^2_{turb}}
\end{equation}
and
\begin{equation}
\sigma_0 = \frac{8\pi^3 \textrm{k}T_1}{\textrm{h}^2\textrm{c}}
\frac{(l+1)^2}{2l+1}\, \mu^2. 
\end{equation}

In the previous equations, $m_{CO}$ is the CO molecular weight, 
$\nu_0$ is the rest frequency of the molecular transition, $\Delta v$
is the difference between the velocity
$v_{obs}=(\textrm{c}/\nu_0)(\nu-\nu_0)$, corresponding to the 
frequency $\nu$, and the component of the gas velocity along the line  
of sight, $v_k(s)$, $\mu$ is the dipole moment of the CO molecule.
Note that writing Eq.~\ref{eq:phi}, we assumed that the intrinsic
line width depends both on the thermal velocity dispersion in the
gas and on turbulent velocity $v_{turb}$.

As shown in Fig.~\ref{fig:kep_rot} and described in
Sec.~\ref{sec:kine}, the observed velocity patterns in the CO
transitions are in good agreement with the Keplerian rotation of
the disk. We can thus assume that gas moves on circular orbits around
the star characterized by a tangential velocity 
\begin{equation}
v_\phi(r) = \sqrt{ \frac{ \textrm{G} M_{\star}}{r} } ,
\end{equation}   
where $r$ is the radius of the orbit and \Mstar\ is the stellar mass. 
To calculate the velocity $v_k(s)$, component of the tangential
velocity $v_\phi(r)$ along the line of sight, it is useful to define a 
coordinates system centered on the star as shown in
Fig.~\ref{fig:disk}: the $(x,y)$ plane corresponds to the disk
midplane; the observer lies in the $(y,z)$ plane and its 
position it is defined by the inclination $\theta$ and the distance
$d$ from the star; each point of the circumstellar space can be
defined through the cylindrical coordinates $r = \sqrt{x^2+y^2}$,
$\phi=\arctan{(y/x)}$ and $z$. Since $d \gg r$, we can write
\begin{equation}
v_k(s) \cong  v_\phi(r) \cdot \cos\phi \cdot \sin\theta.
\end{equation}    

\begin{figure*}
\includegraphics[angle=270, width=18cm]{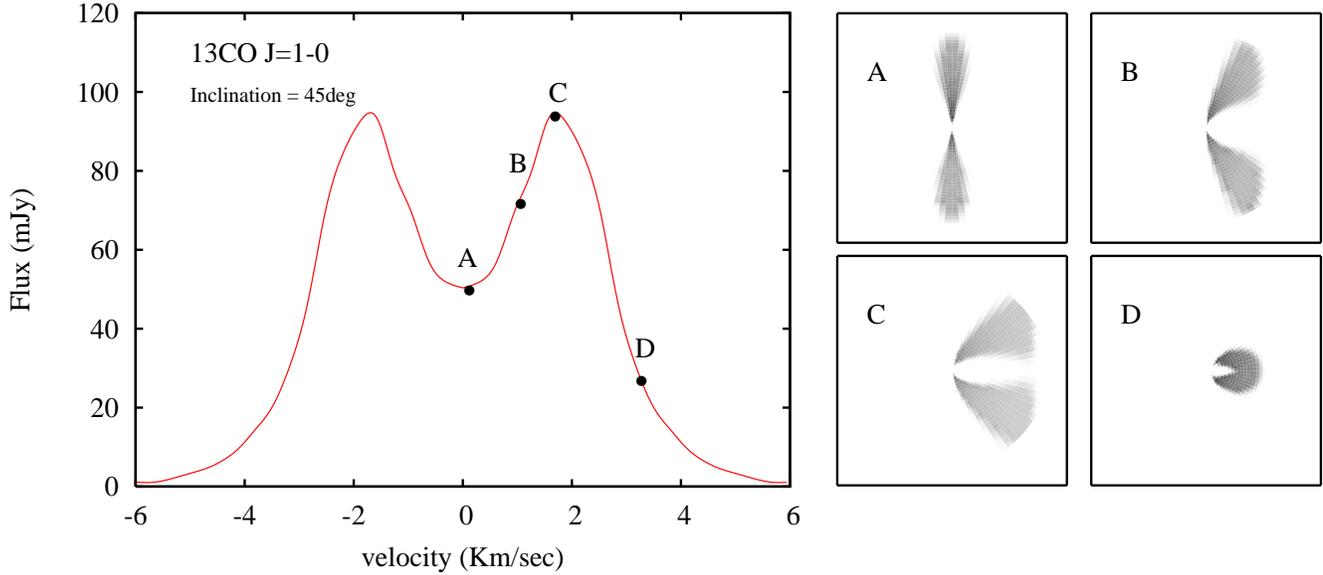}
\caption{\label{fig:mod_map} $^{13}$CO J=1-0 simulated emission for a
  Keplerian disk inclined by 45\grad rotating around a central star of
  1\Msun. The disk outer radius is $R_{out}=100$~AU with a surface
  density $\Sigma(r)\propto r^{-1.5}$. The left panel show the spatial
  integrated line profile. The right panel shows the spatially
  resolved maps at different velocity as labelled in figure.
}
\end{figure*}

Finally, the velocity $v_k(s)$ can be calculated for each direction,
knowing the geometrical transformations between the coordinate $s$
along the line of sight and the cylindrical coordinates $r$ and $\phi$.

In order to solve the described set of equations, we thus need an
expression for the circumstellar mass density $\rho(s)$ and the
temperature of the emitting gas $T_{CO}(s)$. In both cases we can
assume the cylindrical symmetry and write $\rho(s) \equiv \rho(r,z)$
and $T_{CO}(s) \equiv T_{CO}(r)$. 

For the emitting gas temperature, we choose the parameterization
\begin{equation}
T_{CO}(r) = T_{CO}(r_0)(r/r_0)^{-q},
\end{equation}
while the mass density is calculated assuming that the disk is in
hydrostatic equilibrium and vertically isothermal in the inner region 
at the midplane temperature $T_m(r)$, through the relation 
\begin{equation}
\rho(r,z)= \rho_0(r) \cdot e^{-z^2 /2h^2 (r)},
\end{equation}   
valid between the disk inner and outer radii $R_{in}$ and $R_{out}$.
The density on the disk midplane $\rho_0(r)$ can by expressed as
function of the surface mass density $\Sigma(r) = \Sigma_0
(r/r_0)^{-p}$ through the relation 
\begin{equation}
\label{eq:rho_0}
\rho_0(r) = \frac{\Sigma(r)}{\sqrt{2\pi} h(r)}.
\end{equation}
Finally, the pressure scale $h(r)$ is given by the relation
\begin{equation}
h(r) = \sqrt{ \frac{2r^3\textrm{k}T_i(r)}{G M_{\star} m_0} },
\end{equation} 
where $T_i(r)$ is obtained solving the structure of a
stellar-irradiated passive disk as described in Dullemond et
al.~(2001).  

Note that the emitting gas temperature $T_{CO}$ has been parametrized
independently of the disk interior temperature $T_i$, which governs
the density structure of the disk. As pointed out by Dartois et
al. (\cite{Dea03}), the more optically thick CO transitions may by
good tracer of the disk surface where the gas temperature is different
from the disk interior. 

The resulting model, produces brightness maps for each frequency (see
Fig.\ref{fig:mod_map})  that can be compared with the observations
presented in  Fig.~\ref{fig:12CO} and \ref{fig:13CO}.


\end{appendix}

\end{document}